\documentstyle[prl,aps,preprint,tighten,epsf]{revtex}

\def\half{\frac{1}{2}}
\input BoxedEPS
\SetRokickiEPSFSpecial  
\HideDisplacementBoxes
\begin{document}
\draft


\preprint{\vbox{Submitted to Physics Letters B\null\hfill\rm
    MIT-CTP-2711}}
\title{Finite Quantum Fluctuations About Static Field Configurations}

\author{E.~Farhi$^{\rm a}$, N.~Graham$^{\rm a}$, P.~Haagensen$^{\rm
    a}$, and R.~L.~Jaffe$^{\rm a,b}$ }

\address{{~}\\$^{\rm a}$Center for Theoretical Physics, Laboratory for
  Nuclear Science
  and Department of Physics \\
  Massachusetts Institute of Technology,
  Cambridge, Massachusetts 02139 \\
  and \\
  $^{\rm b}$RIKEN BNL Research Center \\
  Brookhaven National Laboratory, Upton, NY 11973\\
  {~}}

\maketitle

\begin{abstract}

        We develop an unambiguous and practical method to calculate
        one-loop quantum corrections to the energies of classical
        time-independent field configurations in renormalizable
        field theories.  We show that the standard perturbative
        renormalization procedure suffices here as well.  We apply our
        method to a simplified model where a charged scalar couples to a
        neutral ``Higgs" field, and compare our results to the
        derivative expansion.
\end{abstract}

 \narrowtext

\section{Introduction}

Solitons arise in many model field theories\cite{Cole} where
nontrivial time-independent solutions to the classical field equations
exist.  The fluctuations of quantum fields about classical
configurations are difficult to study and can qualitatively affect,
even destabilize, solitons.  Topological arguments support the
stability of some particularly interesting solitons. However, much less is
known about the fate of ``non-topological solitons'' that appear as
minima in the classical action, but have no deeper claim to stability
when quantum effects are taken into account.

The quantum corrections to the energies of classical field configurations 
are
typically highly divergent in $3+1$ dimensions.  In some cases, like
the Skyrme model\cite{Sk}, the underlying theory is
non-renormalizable, so the quantum
contribution to the soliton's energy is unavoidably cut-off dependent
and ambiguous, like any other radiative correction in
a non-renormalizable theory.

In this Letter we describe a systematic and efficient procedure for
calculating the quantum fluctuations about time-independent field
configurations in renormalizable field theories. 
We show that all divergences can be removed by the
same renormalization procedure that renders the perturbative sector of
the theory finite.  The only ``ambiguities'' are the well known scheme
and scale dependences of the renormalization prescription that are
resolved completely in the perturbative sector.  The result of our
program is a renormalized quantum ``effective energy'',
whose 
non-trivial minima (if they exist) describe solitons in the 
quantum theory.

First we show how to regulate and renormalize the divergences in the
sum over quantum fluctuations. Then we develop calculational methods
that are efficient and practical enough that quantum effects can be
included in a search for stable field configurations. Our results
take the form of an effective energy, ${\cal E}(\phi(\vec x),m,\{g\})$,
depending on the ``profile function'' of the renormalized field,
$\phi(\vec x)$, the renormalized mass, $m$, and various renormalized 
couplings,
$\{g\}$, defined in the usual perturbative sector of the model. One may
then search over the parameter space characterizing $\phi(\vec x)$ for
minima of ${\cal E}$ while holding $m$ and $\{g\}$ fixed. Here we
treat the simple case of a charged scalar coupled to the field $\phi$.
Our methods can be generalized straightforwardly to models including
fermions, gauge fields and self-coupled scalars.  However, our
approach is limited in that we work only to order $\hbar$, and we 
only consider spherically symmetric profile functions.

The possibility that the top quark in the standard model might be
described as a non-topological soliton \cite{Schechter}
provided the original
motivation for our work. The top quark Yukawa coupling, $g_t$, 
leads to its mass, $g_tv$, where $v$ is the
Higgs vacuum expectation value.
For large $g_t$ it
would therefore appear favorable to suppress the Higgs condensate in
the vicinity of the $t$-quark, and the top quark would be a sort of
``bag''.  However, gradient and potential energy terms in the Higgs
sector of the classical action oppose the creation of a such a bag in the
Higgs condensate.  In order to study the problem at the quantum level
it is necessary to regulate and renormalize the divergences in the
$t$-quark fluctuations about a deformation of the Higgs condensate.
It is crucial to hold the renormalized parameters of the standard
model fixed while varying the possible profiles of the Higgs field.

Bagger and Naculich studied this problem by making a derivative
expansion \cite{BN}. They also worked in a large-$N$ approximation,
in which there are no quantum corrections above order $\hbar$.
We would expect that if the $t$-quark is a bag, then the
Higgs field will vary significantly --- $\Delta\phi\sim v$ --- over
distance scales of order the Compton wavelength of the top quark ---
$\lambda \sim 1/g_tv$. However, the $t$-quark mass, $g_tv$,
also sets the scale for the derivative expansion.  
Thus all derivatives are of
the same size, making the expansion unreliable.  Our method
is designed for such situations.

We can trace elements of our approach back to Schwinger's work on QED
in strong fields \cite{Sch}. Schwinger studied the energy of the
electron's quantum fluctuations --- the ``Casimir energy'' --- in the
presence of a prescribed, static configuration of electromagnetic
fields.  He isolated the divergences in low orders of perturbation
theory.  Our work can be viewed as an extension
of Schwinger's to situations where the field is
determined self-consistently by minimizing the total energy of the
system including the Casimir energy.  In addition we complete the
renormalization program and develop practical computational methods in
three dimensions.  Dashen, Hasslacher and Neveu renormalized the
divergent contributions to the energy of the $\phi^4$ kink and  
sine-Gordon soliton in 1+1
dimensions using a simple version of the method we propose
here \cite{DHN}. Ambiguities in these models recently pointed out
and studied by Rebhan and van Nieuwenhuizen \cite{vN} can also be
resolved with our methods.
Studies of solitons in renormalizable
models often note that the divergences in the quantum contribution to
the soliton energy can be cancelled by the available
counterterms \cite{other}.  However, we are not aware of any
work in 3+1 dimensions
in which renormalization of the field configuration
energy is done in a manner
consistent with on-shell mass and coupling constant renormalization in
the perturbative sector.

\section{Formalism}

We consider a renormalizable field theory with a real scalar field $\phi$
coupled to a charged scalar $\psi$. We take the classical potential 
$V(\phi)\propto (\phi^2-v^2)^2$, and
$\psi$ acquires a mass through spontaneous symmetry breaking. 
At the quantum level we put aside the
$\phi$ self-couplings and consider only the effects of the $\phi-\psi$
interactions.  We further restrict ourselves to ${\cal O}(\hbar)$
effects in the quantum theory, which correspond to one-loop diagrams.

Our model is defined by the classical action
\begin{eqnarray}
  S[\phi,\psi]&=&\int d^4x\left\{ \half
  (\partial_\mu\phi)^2-\frac{\lambda}{4!}(\phi^2-v^2)^2 +
  \partial_\mu\psi^\ast\partial^\mu\psi - g\psi^\ast\phi^2\psi\right.
 \\ \nonumber
  &&\ \ \left. + a (\partial_\mu\phi)^2 - b (\phi^2-v^2) - c (\phi^2-v^2)^2 
\phantom{\half}\right\}\ ,
\label{I.1}
\end{eqnarray}
where we have separated out the three counterterms necessary for
renormalization and written them in a convenient form.
  At one-loop order in $\psi$, these are the only
counterterms required.

We quantize around the classical vacuum $\phi=v$ and define
$h=\phi-v$, so that
\begin{eqnarray}
\label{I.1a}
  S[h,\psi] &=& \int d^4x\left\{\half(\partial_\mu h)^2 - \frac{m^2}{8v^2}
  (h^2 + 2vh)^2 + \partial_\mu\psi^\ast\partial^\mu\psi -
  M^2\psi^\ast\psi - g(h^2+2vh)\psi^\ast \psi\right. \\ \nonumber
  &&\ \ \left. + a(\partial_\mu h)^2 - b(h^2+2hv) - c(h^2+2hv)^2 
\phantom{\half}\right\}
\end{eqnarray}
where $M=\sqrt{g}v$ is the $\psi$ mass and $m^2 = \lambda v^2/3$
is the $h$ mass.

The one-loop quantum effective action for $h$ is obtained by 
integrating out $\psi$ to leading order in $\hbar$.
We are interested in time-independent field configurations
$h=h(\vec x)$, for which the effective action yields an
effective energy ${\cal E}[h]$ that has three parts:
\begin{equation}
  {\cal E}[h] = {\cal E}_{\rm cl}[h] + {\cal E}_{\rm ct}[h] + {\cal
    E}_{\psi}[h]\ ,
\end{equation}
where ${\cal E}_{\rm cl}[h]$ is the classical energy of $h$,
\begin{equation}
  {\cal E}_{\rm cl}[h] = \int d^3x\ \left\{ \half |\vec\nabla h|^2 +
  \frac{m^2}{8v^2} (h^2 + 2vh)^2 \right\}\ ,
\label{I.1b}
\end{equation} 
${\cal E}_{\rm ct}[h]$ is the counterterm contribution,
\begin{equation}
  {\cal E}_{\rm ct}[h] = \int d^3x\left\{ a |\vec\nabla h|^2 +
  b(h^2+2hv) + c(h^2+2hv)^2 \right\}\ ,
\label{I.1b1}
\end{equation}
and ${\cal E}_{\psi}[h]$ is the one-loop quantum contribution from
$\psi$.  ${\cal E}_{\rm ct}[h]$ and ${\cal E}_{\psi}[h]$
are divergent, but we will see explicitly that these divergences
cancel for any configuration $h(\vec x)$.

\begin{figure}
$$
\BoxedEPSF{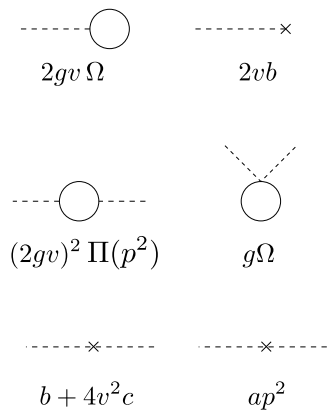 scaled 500}  
$$
\caption{One-loop diagrams.}
\label{figure1}
\end{figure}

We fix the counterterms by applying renormalization conditions
in the perturbative sector of the theory.  Having done so, we have
defined the theory for all $h(\vec x)$.  We choose the on-shell
renormalization conditions
\begin{equation}
  \Sigma_1=0,\quad \Sigma_2(m^2) = 0,\quad {\rm and}\quad
   \left.\frac{d\Sigma_2}{dp^2}\right|_{m^2} = 0,
\label{I.2a}
\end{equation}
where $\Sigma_1$ and $\Sigma_2(p^2)$ are the one- and two-point functions
arising only from the loop and counterterms as seen in
Fig.~\ref{figure1}. 
We denote the one-loop diagrams with one insertion by
$\Omega$ and with two insertions by $\Pi(p^2)$, and find
\begin{eqnarray}
  \Sigma_1 &= &2vg\Omega + 2vb\ ,\nonumber\\
  \Sigma_2(p^2) & = &(2vg)^2 \Pi(p^2) +g\Omega + b + (2v)^2c + a p^2\ .
\label{I.2b}
\end{eqnarray}
Defining
\begin{equation}
\Pi'(p^2) \equiv\frac{d\Pi(p^2)}{dp^2}\ ,
\end{equation}
the renormalization conditions eq.(\ref{I.2a}) then yield
\begin{equation}
a = -(2vg)^2 \Pi'(m^2)\ ,\ b = -g\Omega\ ,\ c = g^2(m^2 \Pi'(m^2)- 
\Pi(m^2))\ ,
\end{equation}
which we then substitute into the counterterm energy, eq.~(\ref{I.1b1}).

Now we consider the calculation of ${\cal E}_{\psi}[ h]$. This energy
is the sum over zero point energies, $\half\hbar\omega$, of the modes of
$\psi$ in the presence of $ h(\vec x)$,
\begin{equation}
  {\cal E}_\psi[ h] = \sum_\alpha E_\alpha[ h]
\label{I.3}
\end{equation}
where $E_\alpha$ are the positive square roots of the eigenvalues of a
single particle Hamiltonian, $\hat H$, given by
\begin{equation}
  \hat H = -\vec\nabla^2 + M^2 + g(h^2 + 2vh).
\label{I.4}
\end{equation}
The fact that $\psi$ is complex accounts for the absence of $\half$
in eq.~(\ref{I.3}).

${\cal E}_\psi$ is highly divergent.  However our model is renormalizable and
therefore the counter\-terms fixed in the presence of the trivial $ h$ {\it must\/}
cancel all divergences in ${\cal
  E}_\psi$.  Rather than attempt to regulate the divergent sum in
eq.~(\ref{I.3}) directly, we study the density of states that defines the
sum. We can 
isolate the terms that lead to 
divergences in ${\cal E}_\psi$ and renormalize them using conventional 
methods. 

For fixed $h(\vec x)$ the spectrum of
$\hat H$ given in eq.~(\ref{I.4}) consists of a finite 
number (possibly zero) of
normalizable bound states and a continuum beginning at $M^2$,
parameterized by 
$k$, with $E(k)=+\sqrt{k^2+M^2}$. Furthermore, $\hat H$ depends on $h$ only
through the combination
\begin{equation}
\chi = h^2+2hv\ ,
\end{equation}
so we can consider ${\cal E}_\psi$ to be a functional
of $\chi$.  We restrict ourselves to spherically symmetric $h$.  Then
\begin{equation}
  {\cal E}_\psi[\chi] = \sum_j E_j + \sum_\ell (2\ell +1) 
\int dk \rho_\ell(k) E(k)
\end{equation}
where $\rho_\ell(k)$ is the density of states in $k$ in the $\ell^{\rm th}$
partial wave and the $E_j$ are the bound state energies. $\rho_\ell(k)$ 
is finite, but the sum over $\ell$ and the integral over $k$ are 
divergent. Furthermore 
\begin{equation}
  \rho_\ell(k) = \rho_\ell^{\rm free}(k)
  + \frac{1}{\pi}\frac{d\delta_\ell(k)}{dk}\ ,
    \label{I.6}
\end{equation}
where $\delta_\ell(k)$ is the usual scattering phase shift for the
$\ell^{\rm th}$ partial wave, and $\rho_\ell^{\rm free}(k)$
is the free ($g=0$) density of states. This
relationship between the density of states and the derivative of the
phase shift is shown for example in \cite{Sch}.

At the outset, we subtract $\rho^{\rm free}(k)$ from the density 
of states since we wish to compare ${\cal E}_\psi[\chi]$ to ${\cal E}_\psi[0]$.
Viewing ${\cal E}_{\psi}[\chi]$ as the sum of one loop
diagrams, we see that only the diagrams with one or two insertions 
of $g\chi$ are divergent. A diagram with $n$ insertions corresponds to
the $n^{\rm th}$  term in the Born expansion, so all possible
divergences can be eliminated by subtracting the first and second Born
approximations from the phase shifts that determine the
density of states. Standard methods allow us to construct the 
Born approximation for the phase shifts \cite{Schiff}, which is a
power series in the ``potential'' $g\chi$.

We define the combination
\begin{equation}
  \bar\delta_\ell(k) \equiv \delta_\ell(k) - \delta^{(1)}_\ell(k) -
  \delta^{(2)}_\ell(k)\ ,
   \label{I.7}
\end{equation}
where $\delta^{(1)}_\ell(k)$ and $\delta^{(2)}_\ell(k)$ are the first
and second Born approximations to $\delta_l(k)$. We then have
\begin{eqnarray}
  {\cal E}_\psi [\chi]&=& \sum_j E_j + \sum_\ell (2\ell+1) 
\int_0^\infty dk
  \frac{1}{\pi} \frac{d\bar\delta(k)}{dk}E(k) + g\Omega 
  \int \frac{d^3\! p}{(2\pi)^3} \tilde \chi(\vec p) \\ \nonumber
  &&\ + g^2 \int \frac{d^3\! p}{(2\pi)^3} \Pi(-\vec p\;\mbox{}^2)
  |\tilde\chi(\vec p)|^2
  \label{I.7a}
\end{eqnarray}
where 
\begin{equation}
\tilde \chi(\vec p) =\int d^3x \chi(\vec x) e^{-i\vec p \cdot \vec x}\ ,
\end{equation}
and likewise for $\tilde h(\vec p)$. Both  $\tilde h$ and 
$\tilde \chi$ are real and
depend only on $q\equiv|\vec p|$ for spherically symmetric $h$.
We have subtracted out the order $g$ and $g^2$ contributions by using
$\bar \delta_\ell(k)$ instead of $\delta_\ell(k)$, and added them back
in by using their explicit diagrammatic representation in terms of the
divergent constant $\Omega$ and the divergent function $\Pi(p^2)$.

We can now combine ${\cal E_\psi}$ and ${\cal E_{\rm ct}}$ and obtain a
finite result:
\begin{eqnarray}
  {\cal E_\psi} + {\cal E_{\rm ct}} &=& \sum_j E_j + \sum_\ell (2\ell+1)
  \int_0^\infty dk \frac{1}{\pi} \frac{d\bar\delta_{\ell}(k)}{dk}E(k)
  +\Gamma_2[h]
   \label{I.8}
\end{eqnarray}
where 
\begin{equation}
  \Gamma_2[h] =
  g^2 \int \frac{q^2dq}{2\pi^2} \left[\left(\Pi(-q^2) 
  - \Pi(m^2) + m^2 \Pi'(m^2)\right) \tilde \chi(q)^2
  +4v^2q^2 \Pi'(-q^2) \tilde h(q)^2 \right] \ .
   \label{I.8a}
\end{equation}
$\Pi$ is log divergent, but both 
$\{\Pi(-q^2)
- \Pi(m^2)\}$ and $\Pi'$ are finite, so $\Gamma_2[h]$ is finite as well.

Each term in the Born approximation to the phase shift goes to zero at
$k=0$, so by Levinson's theorem $\bar\delta_\ell(0) = \delta_\ell(0) =
\pi n_\ell$ where $n_\ell$ is the number of bound states with 
angular momentum $\ell$.  As $k\to\infty$, $\delta_\ell(k)$ falls off
like $\frac{1}{k}$, $\delta^{(1)}_\ell(k)$ falls off
like $\frac{1}{k}$, and $\delta^{(2)}_\ell(k)$ falls off
like $\frac{1}{k^2}$. Since the Born approximation becomes exact
at large $k$, $\bar\delta_\ell(k)$ falls like $\frac{1}{k^3}$. 
Thus we see that the first subtraction renders each 
integral over $k$ convergent.  The second subtraction makes the
$\ell$-sum convergent. We are then free to integrate by parts in
(\ref{I.8}), obtaining
\begin{equation} 
   {\cal E}[ h]= {\cal E}_{\rm cl}[h]
   +\Gamma_2[ h]
   - \frac{1}{\pi}\sum_\ell (2\ell+1)\int_0^\infty dk\:
   \bar\delta_\ell(k)\frac{k}{E(k)} +\sum_j (E_j-M) \ .
   \label{I.9}
\end{equation}
In this expression 
we see that each bound state contributes its binding energy, $E_j-M$,
so that the energy varies smoothly as we strengthen $h$ and bind more states.

As noted in the Introduction, the representation of the Casimir energy
as a regulated sum/integral over phase shifts plus a limited number of
Feynman graphs was derived by Schwinger for the case of a prescribed
background field.  Our aim is to develop it into a practical tool to
study the stability of non-trivial field configurations $h(r)$.

\section{Calculational Methods}

In this Section we describe the method that allows us to construct
${\cal E}[ h]$ as a functional of $ h$ and search for stationary
points. We now consider the calculation of each of the terms in 
eq.~(\ref{I.9})
in turn.  The classical contribution to the action is evaluated directly by
substitution into eq.~(\ref{I.1b}).
$\Gamma_2[h]$ of eq.~(\ref{I.8a}) is obtained from a Feynman diagram 
calculation,
\begin{eqnarray}
  \Gamma_2 [h]= \frac{g^2}{(4\pi)^2}\int\frac {q^2dq}{2\pi^2}
  \int_0^1 &dx& \left\{ \left[
  \log\frac{M^2+q^2x(1-x)}{M^2-m^2x(1-x)}
  -\frac{m^2 x(1-x)}{M^2-m^2x(1-x)} \right] \tilde \chi (q)^2\right.\\
  \nonumber &&\left. - \frac{x(1-x)}{M^2-m^2x(1-x)}
\ 4v^2q^2\tilde{h}(q)^2
  \right\}\ . \label{II.1}
\end{eqnarray}

The partial wave phase shifts and Born approximations are calculated
as follows.  The radial wave equation is
\begin{equation}
   -u_\ell^{\prime\prime} +\left[\frac{\ell(\ell+1)}{r^2}
   +g\chi(r)\right]u_\ell=k^2 u_\ell, \
\label{II.2}
\end{equation}
where $k^2>0$, and $\chi(r)\to 0$ as $r\to \infty$.
We introduce two linearly independent solutions to
eq.~(\ref{II.2}), $u_\ell^{(1)}(r)$ and $u_\ell^{(2)}(r)$, defined by
\begin{eqnarray}
   u_\ell^{(1)}(r) &=& e^{i\beta_\ell(k,r)}r h_\ell^{(1)}(kr) \\ 
\nonumber
   u_\ell^{(2)}(r) &=& e^{-i\beta_\ell^\ast(k,r)}r h_\ell^{(2)}(kr) 
\label{II.3}
\end{eqnarray}
where $h_\ell^{(1)}$ is the spherical
H\"ankel function asymptotic to $e^{ikr}/r$ as $r\rightarrow\infty$,
$h_\ell^{(2)}(kr) = h_\ell^{(1)\ast}(kr)$, and  
$\beta_\ell(k,r) \to 0$ as $r \to \infty$, so that 
$u_\ell^{(1)}(r)\to e^{ikr}$ and
$u_\ell^{(2)}(r)\to e^{-ikr}$ as $r\to\infty$.  The scattering
solution is then
\begin{equation}
u_{\ell}(r) = u_\ell^{(2)}(r) + e^{2i\delta_\ell(k)} u_\ell^{(1)}(r)\ ,
\end{equation}
and obeys $u_{\ell}(0) = 0$.  Thus we obtain
\begin{equation}
   \delta_\ell(k)=2\:{\rm Re}\:\beta_\ell(k,0).
\label{II.4}
\end{equation}
Furthermore, $\beta_\ell$ obeys a
simple, non-linear differential equation obtained by substituting
$u^{(1)}_\ell$ into eq.~(\ref{II.2}),
\begin{equation}
   -i\beta_\ell^{\prime \prime}
   -2ikp_\ell(kr)\beta_\ell^\prime
   +2(\beta_\ell^\prime)^2+\half g\chi(r) = 0\ ,
\label{II.5}
\end{equation}
where primes denote differentiation with respect to $r$, and
\begin{equation}
        p_\ell(x)=\frac{d}{dx} \ln \left[ xh_\ell^{(1)}(x)\right]
\label{II.6}
\end{equation}
 is a simple rational function of $x$.

We solve eq.~(\ref{II.5}) numerically, integrating from $r=\infty$ 
to $r=0$
with $\beta_\ell^\prime(k,\infty)=\beta_\ell(k,\infty)=0$, to get the 
exact phase
shifts.  To get the Born approximation to $\beta_\ell$, we solve the 
equation
iteratively, writing $\beta_\ell = g\beta_{\ell 1} + g^2\beta_{\ell 2} +
\dots$, where $\beta_{\ell 1}$ satisfies
\begin{equation}
   -i\beta_{\ell 1}^{\prime \prime}
   -2ikp_\ell(kr)\beta_{\ell 1}^\prime+\half \chi(r) = 0
\label{II.7}
\end{equation}
and $\beta_{\ell 2}$ satisfies
\begin{equation}
   -i\beta_{\ell 2}^{\prime \prime} 
   -2ikp_\ell(kr)\beta_{\ell 2}^\prime 
   +2(\beta_{\ell 1}^\prime)^2 = 0..
\label{II.8}
\end{equation}
We can solve efficiently for $\beta_{\ell 1}$ and $\beta_{\ell 2}$
simultaneously
by combining these two equations into a coupled differential equation 
for the
vector $(\beta_{\ell 1},~\beta_{\ell 2})$.  This method is much faster than 
calculating the Born terms directly as iterated integrals in $r$ and will
generalize easily to a theory requiring higher-order counterterms.

Having found the phase shifts, we then use Levinson's theorem to
count bound states.  We then find
the energies of these bound states by using a shooting method to
solve the corresponding eigenvalue equation.  We use the effective
range approximation \cite{Schiff} to calculate the phase shift and
bound state energy near the threshold for forming an s-wave bound state.

\section{results}

\begin{figure}
$$
\BoxedEPSF{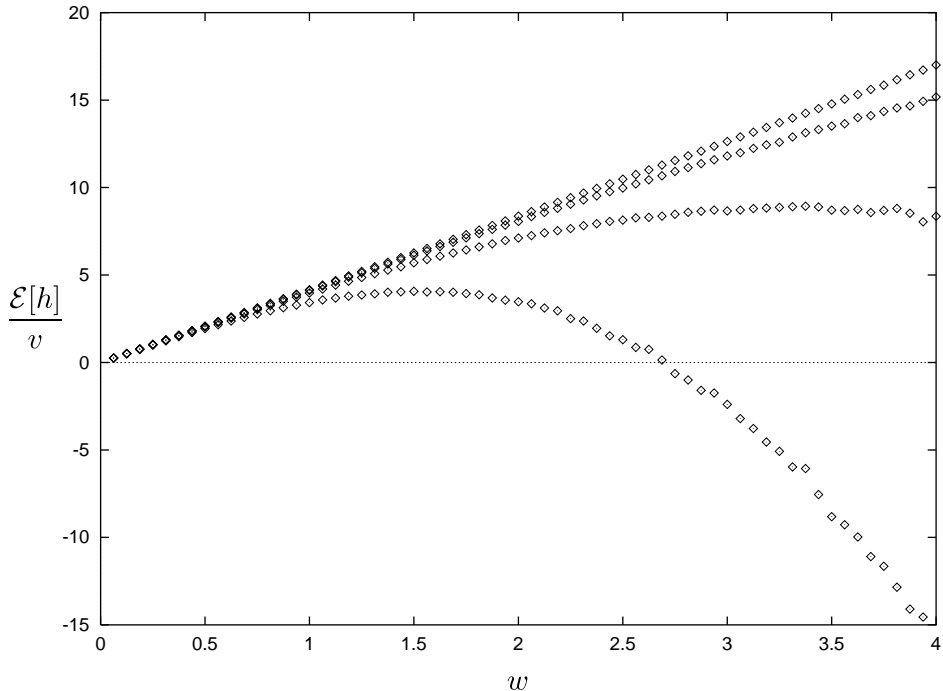 scaled 500}  
$$
\caption{${\cal E}[h]$ in units of $v$, for $d=1$ and 
$g=1,2,4,8$, as a function of $w$.}
\label{figure2}
\end{figure}

For the model at hand, we calculated the energy ${\cal E}[ h]$ for a 
two-parameter ($d$ and $w$) family of gaussian backgrounds
\begin{equation}
h(r)=-dve^{-r^2 v^2/2w^2}\ .
\label{IV.1}
\end{equation}
In Fig.~\ref{figure2}, we show results which are representative of our
findings in general. We plot the energy of configurations with
fixed $d=1$ as a function of $w$, for $g=1,2,4,8$ (from top to bottom).
We note that to this order, for $g=8$ the vacuum is unstable to the 
formation of large $\phi=0$ regions.

To explore whether the charged scalar forms a non-topological soliton
in a given $\phi$ background, we add to ${\cal E}[ h]$ 
the energy of a ``valence'' $\psi$ particle in the lowest bound state.
We then compare this total energy to $M$, the energy of the $\psi$ particle
in a flat background, to see if the soliton is
favored.  This is the scalar model analogue of $t$-quark bag
formation.

For fixed $g$ and $m$, we varied $h$ looking for
bound states with energy $E$ such that $E+{\cal E}[ h]<M$. However,
for those values of $g$ and $m$ where we did find such solutions, we
always found that by increasing $w$, we could make ${\cal E}[ h]<0$,
so that the vacuum is unstable, as we pointed out above in the case 
of $g=8$ in
Fig.~\ref{figure2}.  Thus we find that if
we stay in the $g,m$ parameter region where the vacuum is stable, the
minimum is at $h=0$, so there are no nontopological solitons. 

Although we did not find a non-trivial solution at one-loop order
in this simple model, our calculation demonstrates the practicality
of our method.  We can effectively characterize and search the space
of field configurations, $h(r)$, while holding the renormalized
parameters of the theory fixed.  The same methods can be used
to study solitons in theories with richer structure.
                 
\section{derivative expansion}

Our results are exact to one-loop order.  The derivative expansion, which is
often applied to problems of this sort, should be accurate for slowly varying
$h(r)$.  We found it useful to compare our results with the derivative 
expansion
for two reasons:  first, we can determine the range of validity (in $d$ 
and $w$)
of the derivative expansion; and second, where the derivative expansion is
expected to be valid, it provides a check on the accuracy of our 
numerical work
and C++ programming.  Where expected, the two calculations did agree to the
precision we specified (1 \%).

In our model, the first two terms in the derivative expansion of
the one-loop effective Lagrangian can be calculated to be
\begin{equation}\label{V.1}
{\cal L}_{\rm 1} = {\cal L}_{\rm ct} + \alpha z+\beta z^2
+ {g^2v^4\over 32\pi^2}\left[
(1+z)^2\ln (1+z)-z-{3\over2}z^2\right]+
{g\over48\pi^2v^2}{1\over 1+z} (\partial_\mu z)^2  \ ,
\end{equation}
where $z=g\chi/M^2=(h^2+2hv)/v^2$, $\alpha$ and $\beta$ are
cutoff-dependent constants, and ${\cal L}_{\rm ct}$ is the same
counterterm Lagrangian as we used in Sec. 2.  For $\phi^4$
scalar field theory a similar result was first derived in
\cite{Coleman}. The last term above is proportional to $(\partial
h)^2$, and is completely cancelled by a finite counterterm that
implements the renormalization prescription of Sec. 2. In this
prescription, counterterms also cancel the $\alpha z$ and $\beta z^2$
terms above. Thus the ${\cal O}(p^2)$ derivative expansion for the  effective
energy, to be compared with the phase shift  expression for ${\cal E}[h]$, is

\begin{equation}\label{V.2}
{\cal E}_{\rm DE}[h]=\int d^3\! x\ \left\{
\half (\vec{\nabla}h)^2 +{m^2\over8v^2}(h^2+2hv)^2+
{g^2v^4\over32\pi^2}\left[ (1+z)^2\ln (1+z )-
z-{3\over2}z^2\right]\right\}\ .
\end{equation}

\begin{figure}
$$
\BoxedEPSF{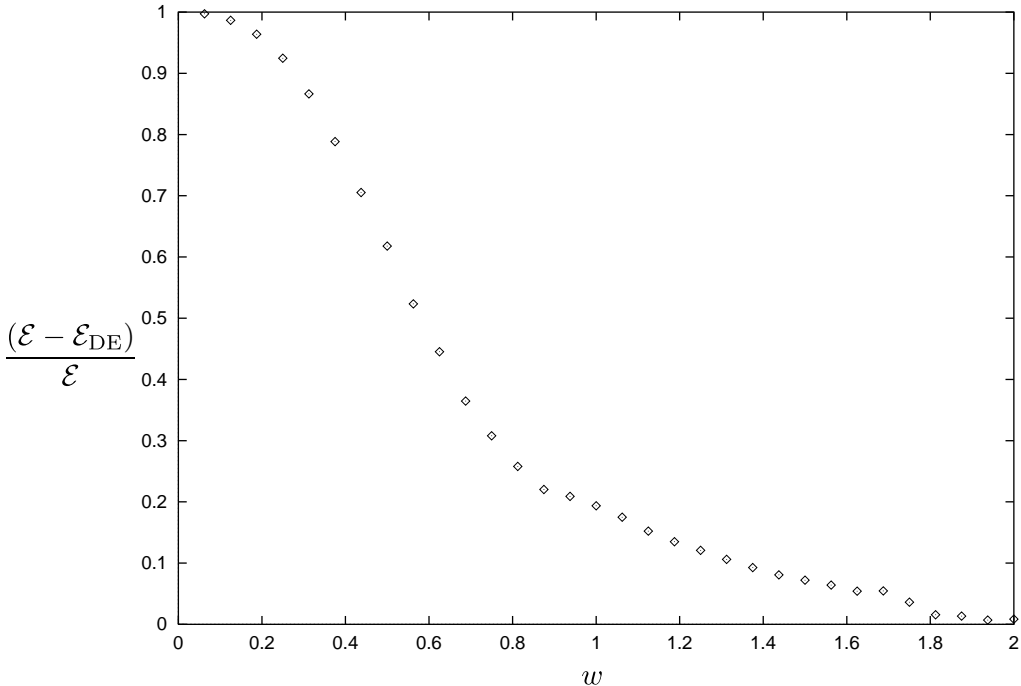 scaled 500}  
$$
\caption{$( {\cal E} -{\cal E}_{\rm DE})/{\cal E}$
for $d=0.25$, $g=4$, as a function of $w$.}
\label{figure3}
\end{figure}

The results of the comparison with the phase shift method can 
be seen in Fig.~\ref{figure3} for $d = 0.25$,
and $g=4$. A similar pattern holds in general for other values of both
$d$ and $g$. As the width becomes larger, the two results merge. This is
as expected, since it is for large widths, and thus small gradients,
that we expect the derivative expansion to yield accurate results. As the
width tends to zero, both results tend to zero, and the fact that the plot 
tends to 1 simply indicates that the derivative expansion result goes to
zero faster than the phase shift result.

\section{conclusions}

We have presented a numerically tractable method for evaluating the one-loop
effective energy of a static background field configuration in a 
renormalizable quantum field theory. Since we rely on calculating
phase shifts, the method is only suitable for rotationally invariant
(or generalized rotationally invariant) backgrounds. The model explored
in the present work is particularly simple and does not support any 
solitons. However, our methods could just as well be applied to any
renormalizable field theory. They can be used to study the one-loop quantum
stability of field configurations in the standard electroweak model as well as
various unified models that support monopoles, strings and the like. 
Topologically non-trivial field configurations with maximal symmetry, like the
``hedgehog solutions'' in chiral models can be studied in this fashion.  
Ultimately we hope to be able to reliably determine whether large Yukawa
couplings may yield solitons in the standard electroweak model.


\vspace*{1cm}

We would like to thank J.~Goldstone, R.~Jackiw, K.~Johnson, K.~Kiers,
S.~Naculich, B.~Scarlet, R.~Schrock, M.~Tytgat, and P.~van Nieuwenhuizen for
helpful conversations, suggestions and references.  This work is
supported in part by funds provided by the U.S.  Department of Energy
(D.O.E.) under cooperative research agreement \#DF-FC02-94ER40818.
and by the RIKEN BNL Research Center.  N.~G. is supported in part by an NSF
Fellowship.



\end{document}